\documentclass[epsf,useAMS,usenatbib,usegraphicx]{mn2e}
\usepackage{epsf}
\usepackage[usenames]{color}

\title[New roAp stars HD\,69013 and HD\,96237 ]{The discovery of rapid 
oscillations in the magnetic Ap stars HD\,69013 and HD\,96237\thanks{Based on 
observations collected at the European Southern Observatory, Chile, as part of 
programmes 080.D-0191(A), 078.D-0192(A) and 078.D-0080(A) }}

\author[Elkin et al.] {V.G.~Elkin$^{1}$, D.W.~Kurtz$^{1}$,
H.L.~Worters$^{2}$, G.~Mathys$^{3}$, B.~Smalley$^{4}$, F.~van Wyk$^{2}$,
\newauthor A.M.S.~Smith$^{4}$ \\
$^{1}$Jeremiah Horrocks Institute of Astrophysics, University of Central
Lancashire, Preston PR1 2HE, UK\\
$^{2}$South African Astronomical Observatory, PO Box 9, Observatory 7935, South 
Africa \\
$^{3}$European Southern Observatory, Casilla 19001, Santiago 19, Chile \\
$^{4}$Astrophysics Group, Keele University, Keele, Staffordshire ST5 5BG\\ }

\begin{document}

\maketitle

\begin{abstract}
We report the detection of short period variations in the stars HD\,69013 and 
HD\,96237. These stars possess large overabundances of rare earth elements and 
global magnetic fields, thus belong to the class of chemically peculiar Ap 
stars of the main sequence. Pulsations were found from analysis of high time 
resolution spectra obtained with the ESO Very Large Telescope using a cross 
correlation method for wide spectral bands, from lines belonging to rare earth 
elements and from the H$\alpha$ core. Pulsation amplitudes reach more than 
200\,m\,s$^{-1}$ for some lines in HD\,69013 with a period of 11.4\,min and about 
100\,m\,s$^{-1}$ in HD\,96237 with periods near 13.6\,min. The pulsations have 
also been detected in photometric observations obtained at the South African 
Astronomical Observatory.
\end{abstract}

\begin{keywords}
Stars: chemically peculiar -- stars: oscillations -- stars: magnetic.
\end{keywords}

\section{Introduction}

The rapidly oscillating Ap (roAp) stars are magnetic main sequence stars that 
pulsate in high radial overtone p~modes with periods in the range of $5.7 - 
21$\,min. They show broad-band photometric amplitudes less than 0.01\,mag, whereas 
rapid radial velocity variations in rare earth element lines can reach several 
km\,s$^{-1}$ (e.g., \citealt{Freyhammer09}; \citealt{Kurtz90}). The roAp stars are 
important targets for the study of the interactions among chemical anomalies, 
magnetic field and pulsations. These stars show abnormal atmospheric structure 
with chemical  stratification (\citealt{Cowley01}; \citealt{Ryab02}). The 
pulsations of these stars also make them promising objects for interior model 
testing using asteroseismology (\citealt{Aerts10}). The roAp stars were discovered 
by \cite{Kurtz82}; at present more than 40 such stars are known.

Interesting and surprising discoveries have been made in recent years, which give 
new insight in the study of pulsating Ap stars. New ground is being broken with 
$\mu$mag precision photometric data from the {\it Kepler} Mission. \cite{BalonaK1} 
discovered a roAp star that pulsates in both high overtone p\,modes and a low-
frequency g\,mode, opening up the possibility of better modelling of the interiors 
of these most peculiar stars.

In 2007 we started a high resolution survey of cool chemically peculiar stars 
based mostly on the photometric catalogue of \cite{Martinez93}. One of the goals 
of the survey is to select stars with high peculiarity and strong magnetic field, 
and to determine their fundamental parameters to select the most promising 
candidates to be roAp stars. Nearly 400 stars have been observed and many of them 
are good roAp candidates for further high time resolution spectroscopic and 
photometric observations. For several of these stars we have obtained such 
observations and here we present the discovery of pulsations for two objects, 
HD\,69013 and HD\,96237.

Both stars were in the list of stars with magnetically split spectral lines found 
by \cite{Freyhammer08b}. HD\,69013 is a typical cool Ap star; HD\,96237 is more 
impressive with significant spectral variability.  The physical parameters $T_{\rm  
eff}$ and $\log g$ indicate that HD\,69013 and HD\,96237 are both main sequence 
stars, situated in the HR diagram where the instability strip crosses the main 
sequence. For many Ap stars in this region of main sequence rapid oscillations 
have been detected, therefore both stars were promising objects for pulsation 
testing.

\section{Observations and data reduction}

High time resolution spectroscopic observations were carried out at the European 
Southern Observatory (ESO) using the Ultraviolet and Visual Echelle Spectrograph 
(UVES) installed at Unit Telescope 2 (UT2) of the Very Large Telescope (VLT). For 
HD\,69013, data were obtained during two high time resolution observing runs on 
2008 January 17 and February 6. For each run 34 spectra were obtained with 
exposure times of 80\,s and readout and overhead times of $\sim21$\,s, 
corresponding to a time resolution of $\sim$101\,s. For HD\,96237 we obtained 34 
spectra on 2008 March 15 with the same exposure and readout times. The wavelength 
region observed is $\lambda\lambda\,4970 - 7010$\,\AA, with a small gap in the 
region around 6000\,\AA\ caused by the space between the two CCDs. The average 
spectral resolution is about $R = 10^5$. The CCD frames were processed using the 
UVES pipeline to extract and merge the echelle orders to 1D spectra that were 
normalised to the continuum.

We also obtained photometric observations of HD\,69013 and HD\,96237 in January, 
February and May 2010. These observations were obtained at the South African 
Astronomical Observatory (SAAO) in Johnson $B$ filter with the 1-m telescope and 
SAAO CCD and STE4 detector and with the modular photometer at the 0.5-m telescope. 
The reduction of photometric observations was done with {\small ESO-MIDAS} 
software and with software developed at SAAO. The lists of the observations are 
shown in Tables\,1 and 2.  

\begin{table}
\centering
\caption{A journal of observations of HD\,69013. The columns give the Julian Date 
(JD) of the start of exposure, the observation time and the number of spectra 
or photometric measurements. }
\label{tab:obs1}
\begin{tabular}{lccl}
\hline
\hline
\multicolumn{1}{c}{JD} & observation  & exposures & Telescope \\
& time (min)   & number & \\
\hline
2454482.546 & 57   & 34   & ESO UT2 VLT       \\
2454502.549 & 57   & 34   & ESO UT2 VLT       \\
2455226.448 & 121  & 362  & SAAO  1-m   \\
2455229.377 & 84   & 252  & SAAO  1-m   \\
\hline
\end{tabular}
\end{table}

\begin{table}
\centering
\caption{A journal of observations of HD\,96237. The columns are the same as
 for Table\,1. }
\label{tab:obs2}
\begin{tabular}{lccl}
\hline
\hline
\multicolumn{1}{c}{JD} & observation & exposures & Telescope \\
& time (min)   & number & \\
\hline
2454540.515 & 57  & 34   & ESO UT2 VLT      \\
2455225.501 & 118 & 354  & SAAO  1-m   \\
2455228.438 & 100 & 301  & SAAO  1-m   \\
2455229.486 & 90  & 541  & SAAO  1-m   \\
2455235.483 & 58  & 184  & SAAO  1-m   \\
2455329.226 & 224 & 1256 & SAAO 0.5-m \\
\hline
\end{tabular}
\end{table}

\section{Pulsation search and analysis}

For roAp stars lines of rare earth elements show higher pulsation amplitudes, 
while the lines of other chemical species, including light elements and iron peak 
elements, show much smaller pulsation amplitude, or show none at all (see, e.g., 
\citealt{Malanushenko98}; \citealt{Kurtz07}). This strange behaviour is explained 
by stratification where rare earth elements concentrate in the upper layers of the 
stellar atmosphere where oscillation amplitudes reach a maximum, while most other 
chemical elements tend to concentrate in deeper layers where the pulsation 
amplitude is lower. Lines of iron peak elements in roAp stars show very low 
pulsation amplitude or none at all (\citealt{Koch01}, \citealt{Elkin08}). 

To search for rapid radial velocity variability we performed cross-correlation of 
sections of the spectrum using {\small ESO-MIDAS} software. We also measured the 
central positions for profiles of individual spectral lines by the centre of 
gravity method. Frequency analyses of radial velocity and photometric time series 
were performed using {\small ESO-MIDAS}'s Time Series Analysis and a discrete 
Fourier transform programme by \citet{kurtz85}.

\subsection{HD\,69013}

\begin{figure}
\centering 
\epsfxsize 8cm \epsfbox{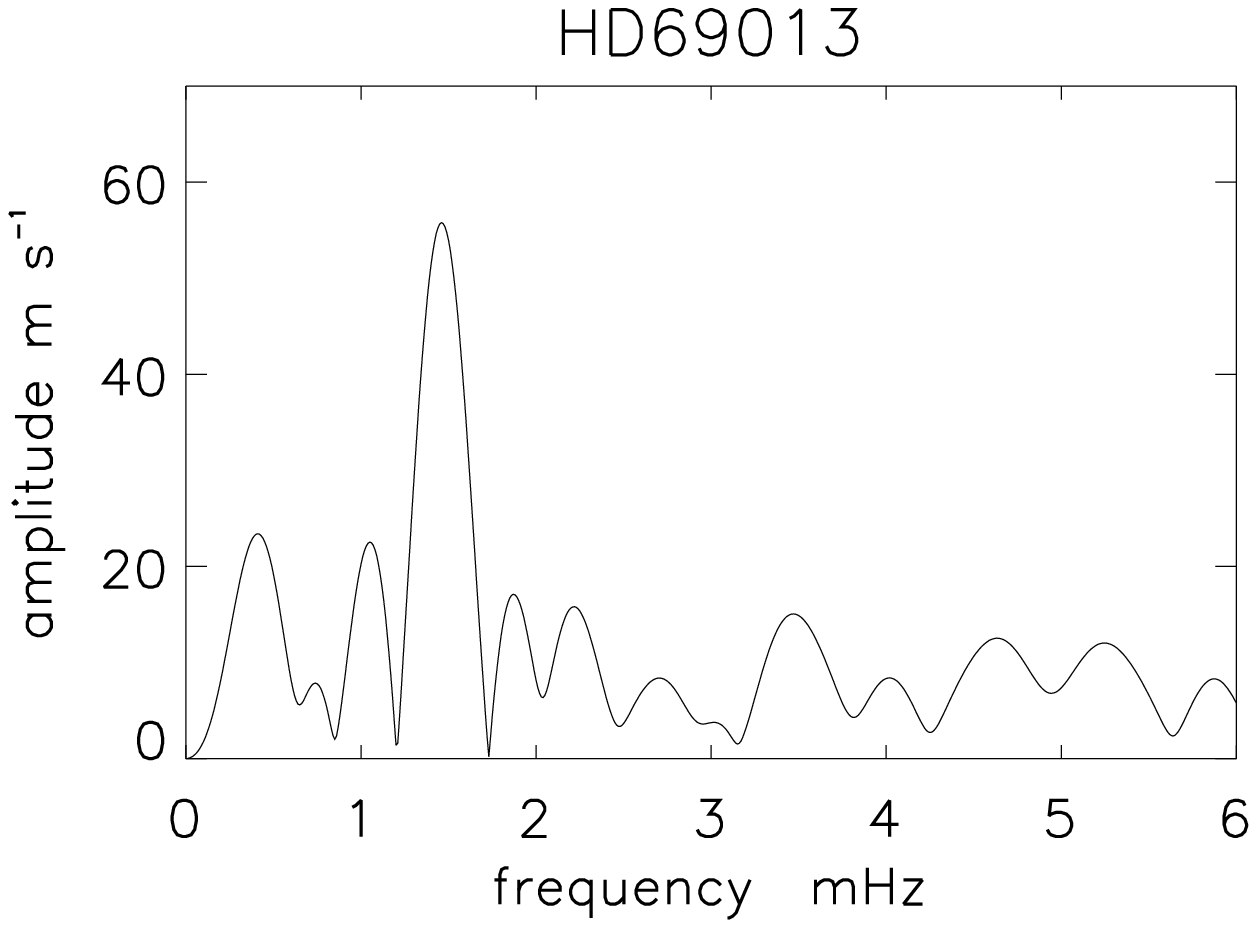}
\epsfxsize 8cm \epsfbox{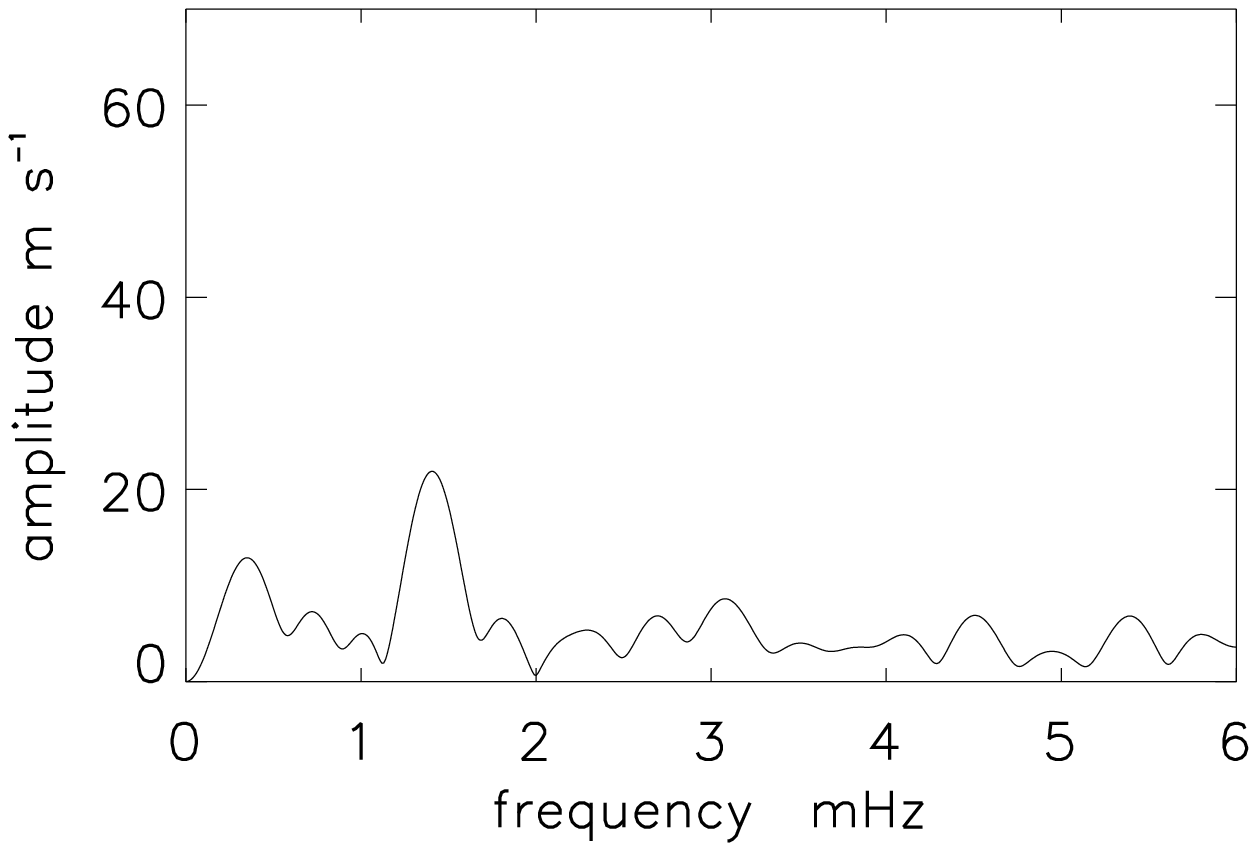}
\caption{\label{69013:freq1} Top: The amplitude spectrum of HD\,69013 obtained 
from cross correlation over the spectral region $5000 - 5800$\,\AA\ for the first 
observing run (see table~1). An average spectrum was used as the template for the 
cross correlation. The frequency peak is at $\nu = 1.46$\,mHz. Bottom: the 
amplitude spectrum for the second spectroscopic run. The frequency peak is $\nu = 
1.42$\,mHz. The pulsation amplitude dropped significantly between the two runs, 
indicating rotational modulation or multiperiodicity.}
\end{figure}

Cross correlation for the spectral band $5000 - 5800$\,\AA\ with an average 
spectrum taken as a template shows obvious rapid oscillations for HD\,69013, as 
can be seen in Fig.\,\ref{69013:freq1}. For the two independent observing data 
sets that we obtained with the ESO VLT telescope we find in the amplitude spectra 
highest peaks at $\nu = 1.46$\,mHz and $\nu = 1.42$\,mHz, correspondingly, with a 
full-width-half-maximum uncertainty of 0.07\,mHz. Hence these two independent 
peaks are at the same frequency within the frequency error.

\begin{figure}
\centering 
\epsfxsize 8cm \epsfbox{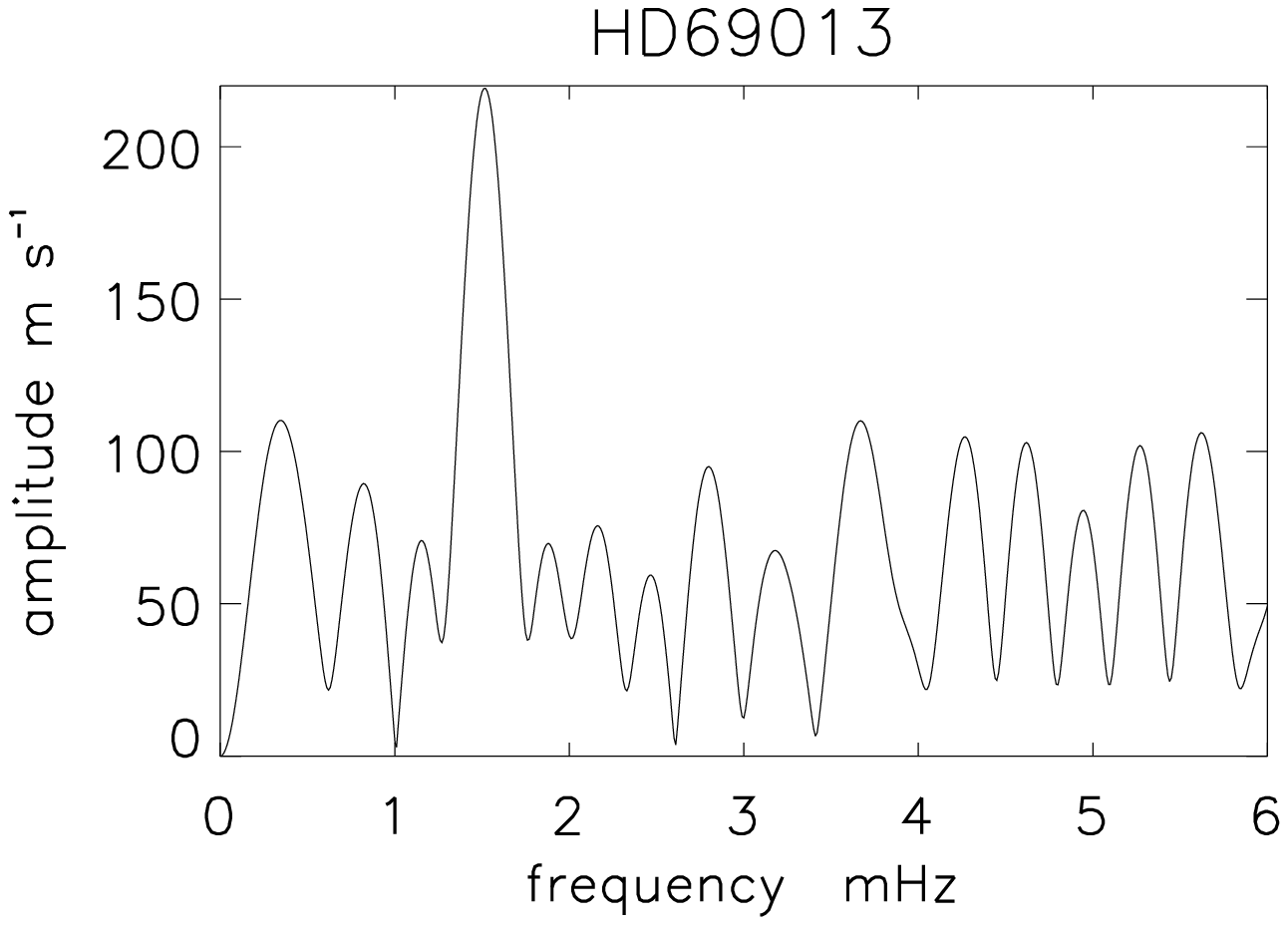}
\epsfxsize 8cm \epsfbox{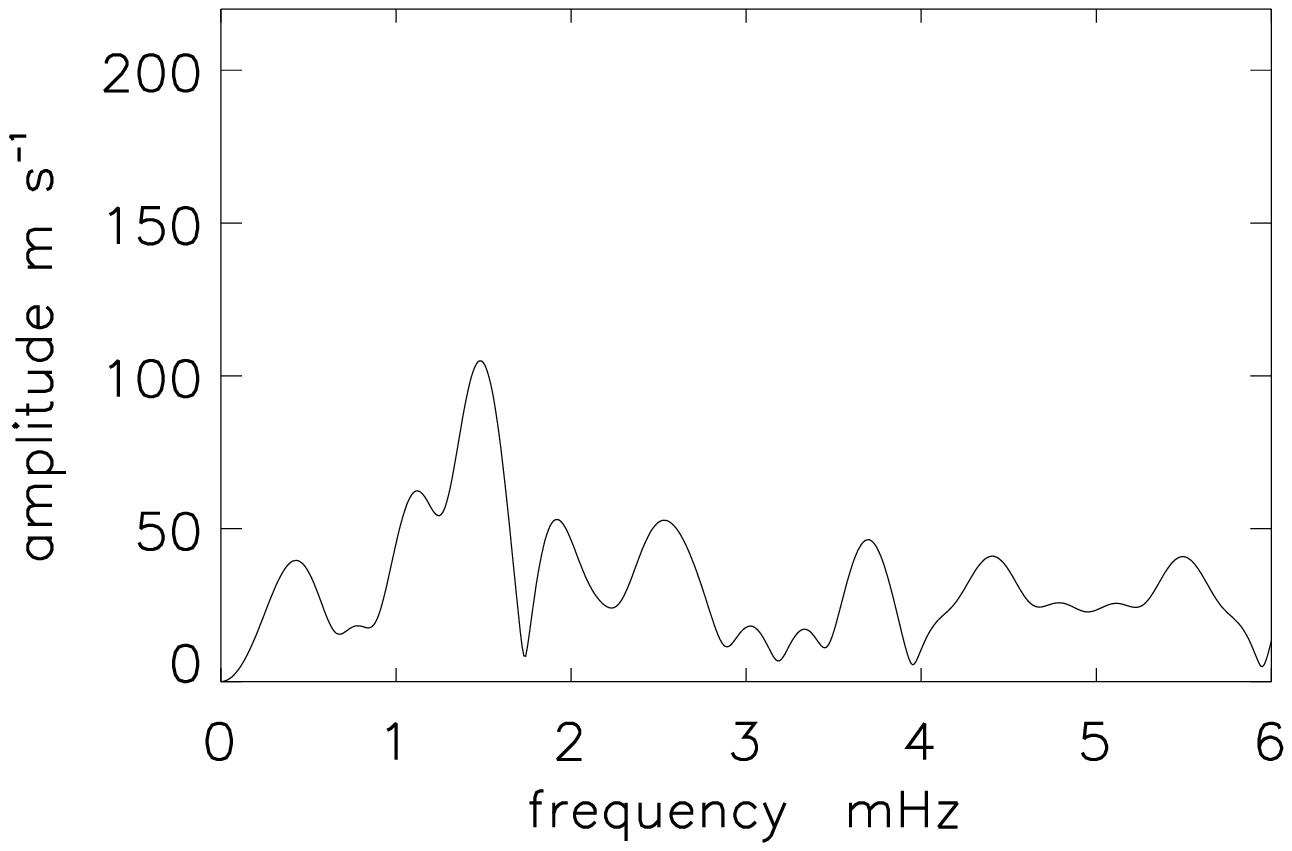}
\caption{\label{69013:priii} Top: the amplitude spectrum of HD\,69013 obtained 
from the line of Pr\,\textsc{iii} at $\lambda$5299.993\,\AA. The frequency peak is 
$\nu = 1.51$\,mHz. Bottom: The amplitude spectrum for the second spectroscopic set 
of HD\,69013 for the same line. The frequency peak is $\nu = 1.48$\,mHz. }
\end{figure}

\begin{figure}
\centering 
\epsfxsize 8cm \epsfbox{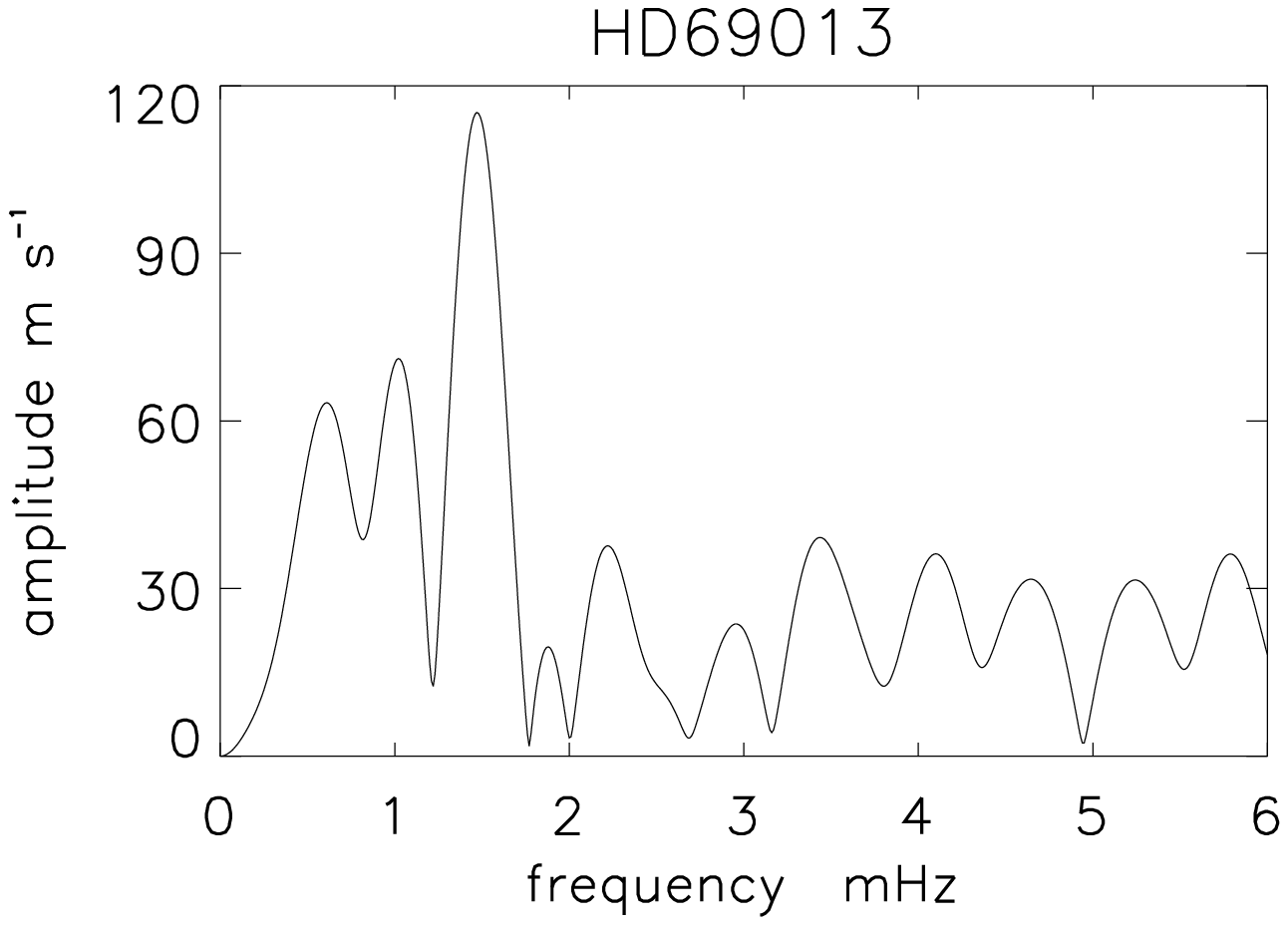}
\epsfxsize 8cm \epsfbox{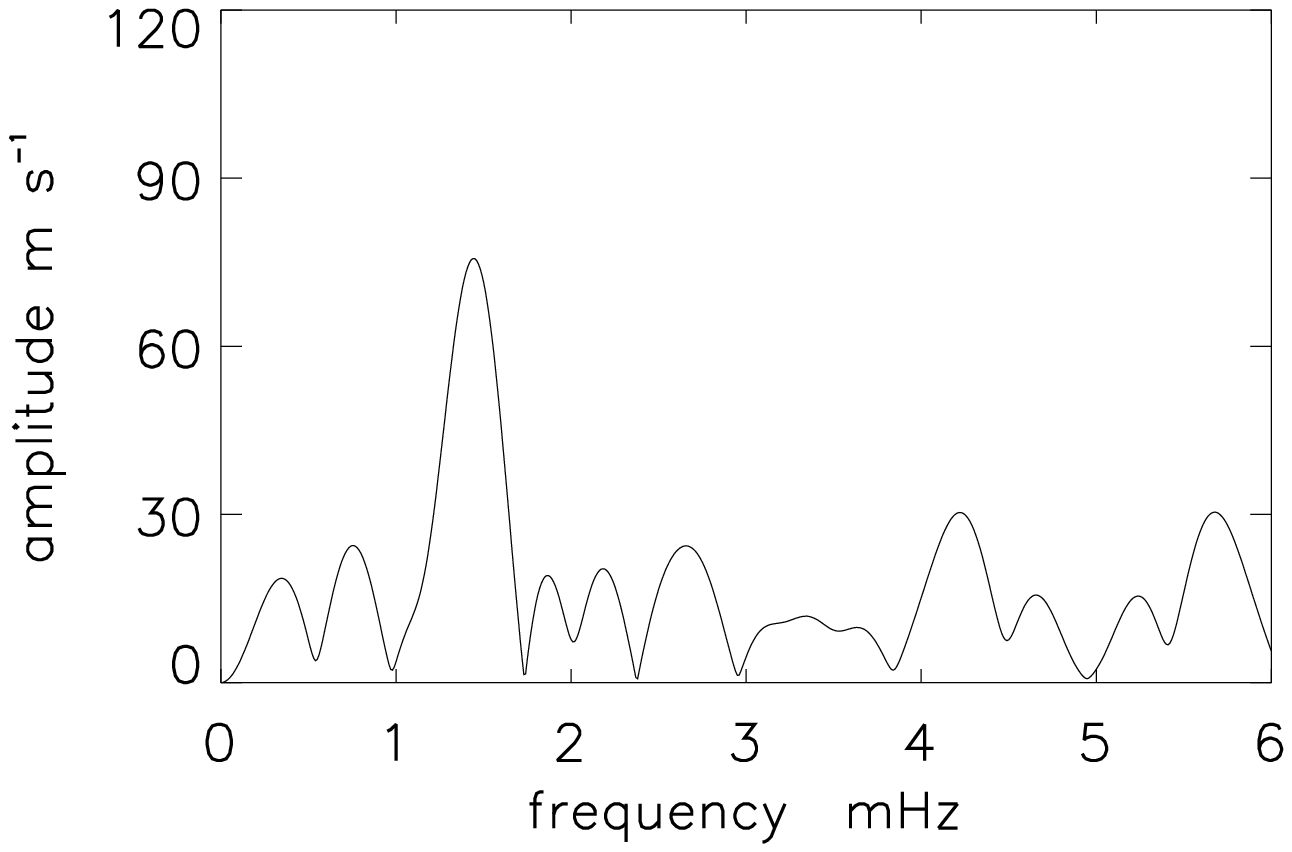}
\caption{\label{69013:ndiii} Top: The amplitude spectrum of HD\,69013 obtained 
from three lines of Nd\,\textsc{iii} ($\lambda\lambda$5050.695\,\AA, 
5102.427\,\AA, 5294.113\,\AA) combined. The frequency peak is $\nu = 1.47$\,mHz. 
Bottom: The amplitude spectrum for the second spectroscopic set of HD\,69013 for 
the same three lines. The frequency peak is $\nu = 1.44$\,mHz. Within the errors 
these two peaks are the same.}
\end{figure}

The spectral lines of the rare earth elements also show pulsation with different 
amplitudes. The highest amplitude we detected was obtained for lines of 
Pr\,\textsc{iii}, shown in Fig.\,\ref{69013:priii}, while lines of 
Nd\,\textsc{iii} reveal a smaller amplitude as shown in Fig.\,\ref{69013:ndiii}. 
The other lines which belong to Eu\,\textsc{ii}, Ce\,\textsc{ii}, La\,\textsc{ii} 
also show pulsations with significant peaks in the amplitude spectra.

The pulsation amplitude is smaller for the second observing run, which suggests 
either rotational modulation or multiperiodicity. While many Ap stars show 
rotational light variations caused by abundance spots usually associated with 
their magnetic poles, there is no such evidence yet found for HD\,69013 
\citep{Freyhammer08b}. This constraint is weak and does not rule out rotational 
modulation pulsation amplitudes. Further observations are needed to study this 
question. 

\begin{figure}
\centering 
\epsfxsize 8cm \epsfbox{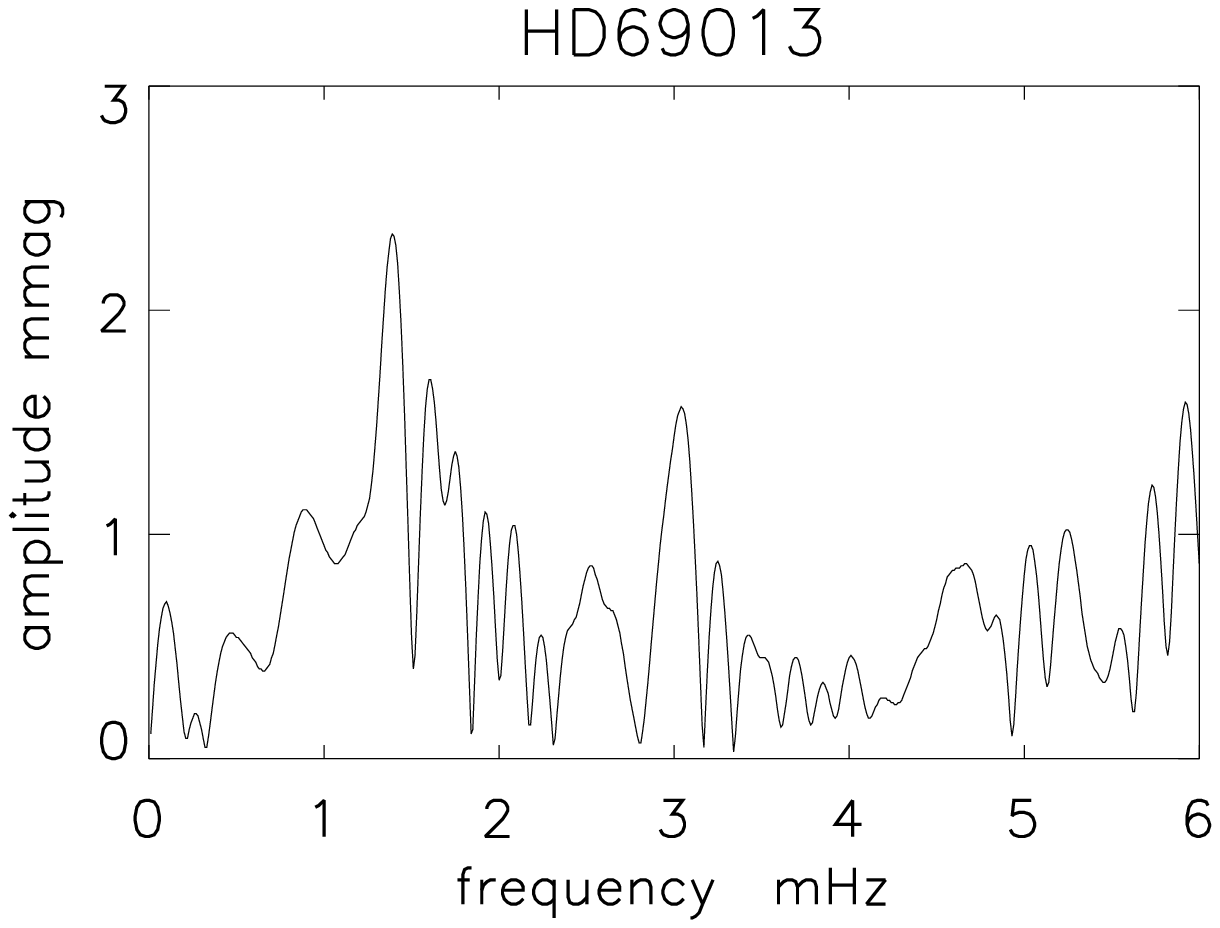}
\epsfxsize 8cm \epsfbox{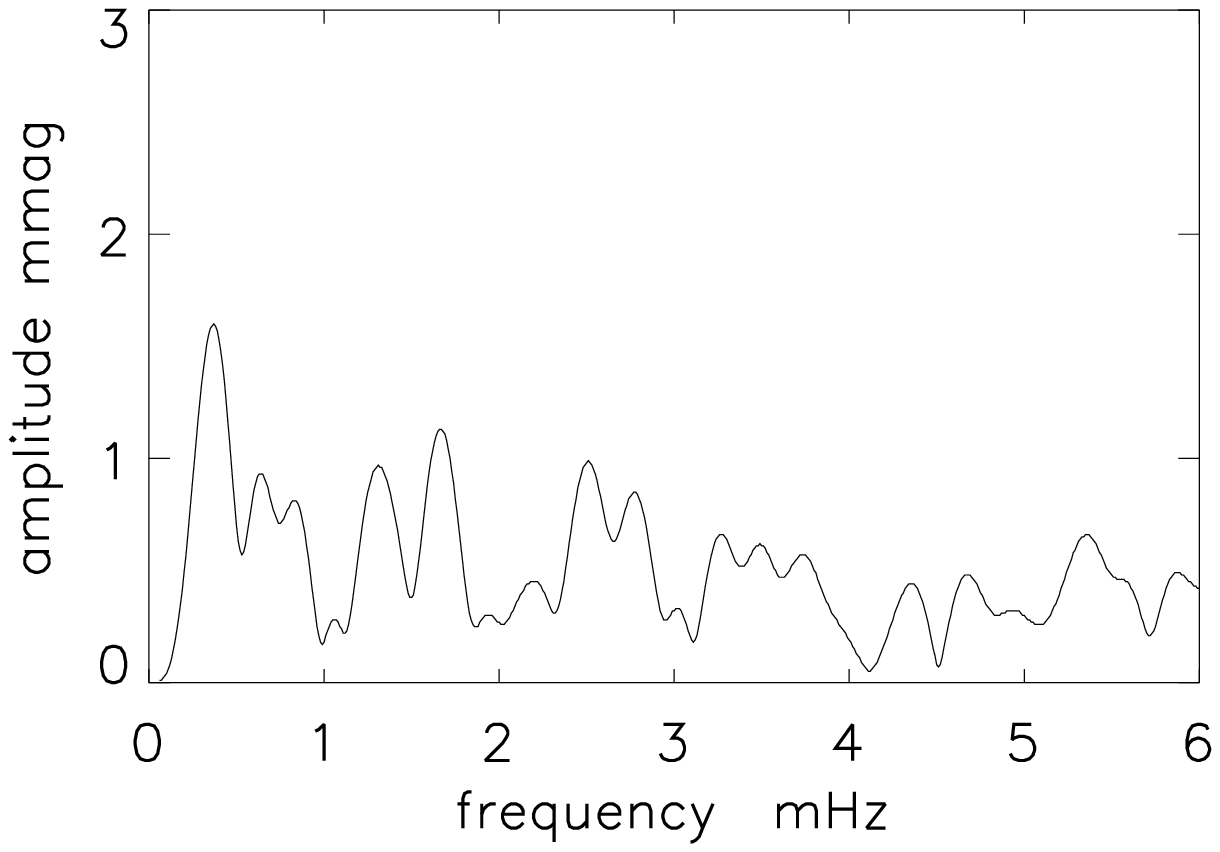}
\caption{\label{69013:saao} Top: The amplitude spectrum of HD\,69013 obtained for 
the first run of SAAO photometry through a $B$ filter. The highest peak has a 
frequency of 1.39\,mHz. Bottom: The second photometric set for the star does not 
show any significant peak for frequencies close to that detected by spectroscopy.}
\end{figure}

The photometric observations of this star obtained at SAAO also show rapid 
oscillations, as can be seen in Fig.\,\ref{69013:saao}. Previous photometric 
observations by  \cite{Nelson93} and \cite{Martinez94} did not detect variations 
in this star. The amplitude spectra of our SAAO photometric observations are shown 
in Fig.\,4. We obtained data on two nights. The first data sets shows a signal at 
the same frequency as for the radial velocity data, but there is no significant 
peak in the second data set, as can be seen in Fig.\,4. 

\subsection{HD\,96237}

\begin{figure}
\centering
\epsfxsize 8cm \epsfbox{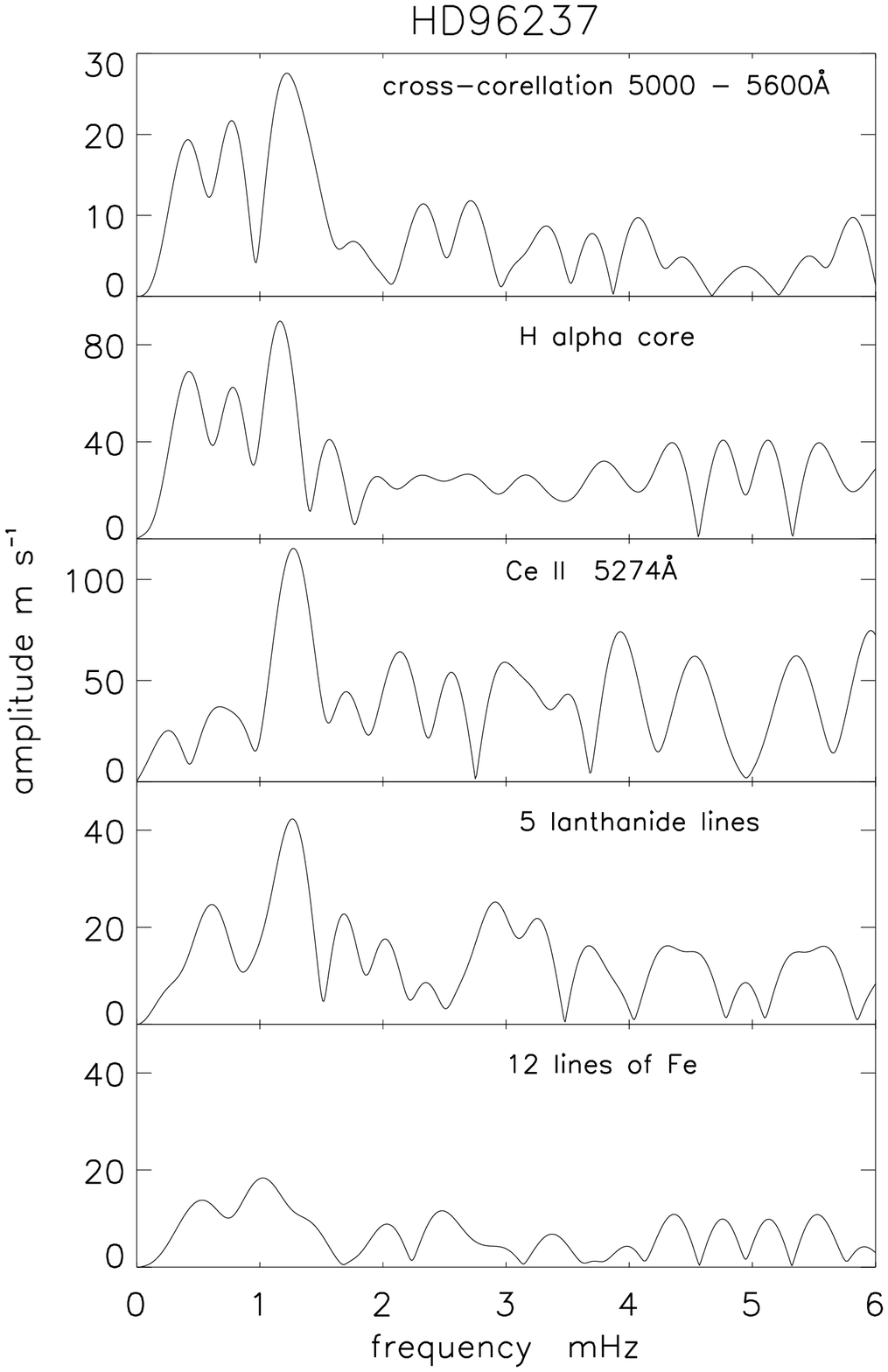}
\caption{\label{96237:ampspec}A cross correlation and Fourier analysis for 34 
spectra of HD\,96237. From top to bottom: the first panel shows a highest peak 
with a frequency of 1.22\,mHz; the second panel has a peak at 1.16\,mHz which is 
in consistent with the first panel within the errors. Both upper panels also show peaks 
with the frequency 0.78\,mHz which is close to that detected in two photometric 
observations(see panels 3 and 4 at Fig.\,\ref{96237:phot}. The peaks in the third 
and forth panels have highest peaks at a frequency of 1.27\,mHz. The fourth panel 
shows an amplitude spectrum for a combination of lines --  Nd\,\textsc{iii} 
5102.43\,\AA, 5294.11\,\AA, Pr\,\textsc{iii} 5284.69\,\AA,  5299.99\,\AA, 
Nd\,\textsc{ii} 5319.81\,\AA. No pulsation signal was detected for a combination 
of 12 lines of iron, as shown at the fifth panel.}
\end{figure}

The case for pulsation in HD\,96237 is not as strong as for HD\,69013, but all of 
the evidence together gives confidence that pulsation has been detected and this 
star is a roAp star. The five panels of Fig.\,\ref{96237:ampspec} show amplitude 
spectra for spectroscopic analysis of HD\,96237. Clear pulsation peaks were 
obtained from cross correlation for the spectral region $5000 - 5600$\,\AA\ using 
an average spectrum as a reference template, and for the core of the H$\alpha$ 
line.

As mentioned above, the spectrum of HD\,96237 is very rich in rare earth element 
lines. We tried to measure pulsations for individual spectral lines which were 
identified by comparison with synthetic spectra calculated with the {\small SYNTH} 
code of \citet{piskunov92}, but found that the pulsation amplitude is too low for 
clear detection in most individual lines, given the noise level in our spectra and 
relatively short observing run. In a few spectral lines of rare earth elements a 
peak corresponding the pulsation frequency detected by cross correlation is 
visible in the amplitude spectrum, as can be seen in the third panel of 
Fig.\,\ref{96237:ampspec} for a single Ce\,\textsc{ii} line.

The combination of several good spectral lines reduces the noise level and 
produces a more reliable picture. The two lower panels in 
Fig.\,\ref{96237:ampspec} show amplitude spectra for combinations of five rare 
earth elements lines giving a clear peak. For twelve lines of Fe\,\textsc{i} and 
Fe\,\textsc{ii} no pulsation is detectable, as is typical of the roAp stars.

As a further test we examined a combination of several telluric lines and found 
that there is no signal above a noise level of 12\,m\,s$^{-1}$. We conclude that 
the case that HD\,96237 is a roAp star is good. Further studies -- particularly at 
the rotation phase where the pulsation amplitude is highest -- will give more 
confidence.

\begin{figure}
\centering 
\epsfxsize 8cm \epsfbox{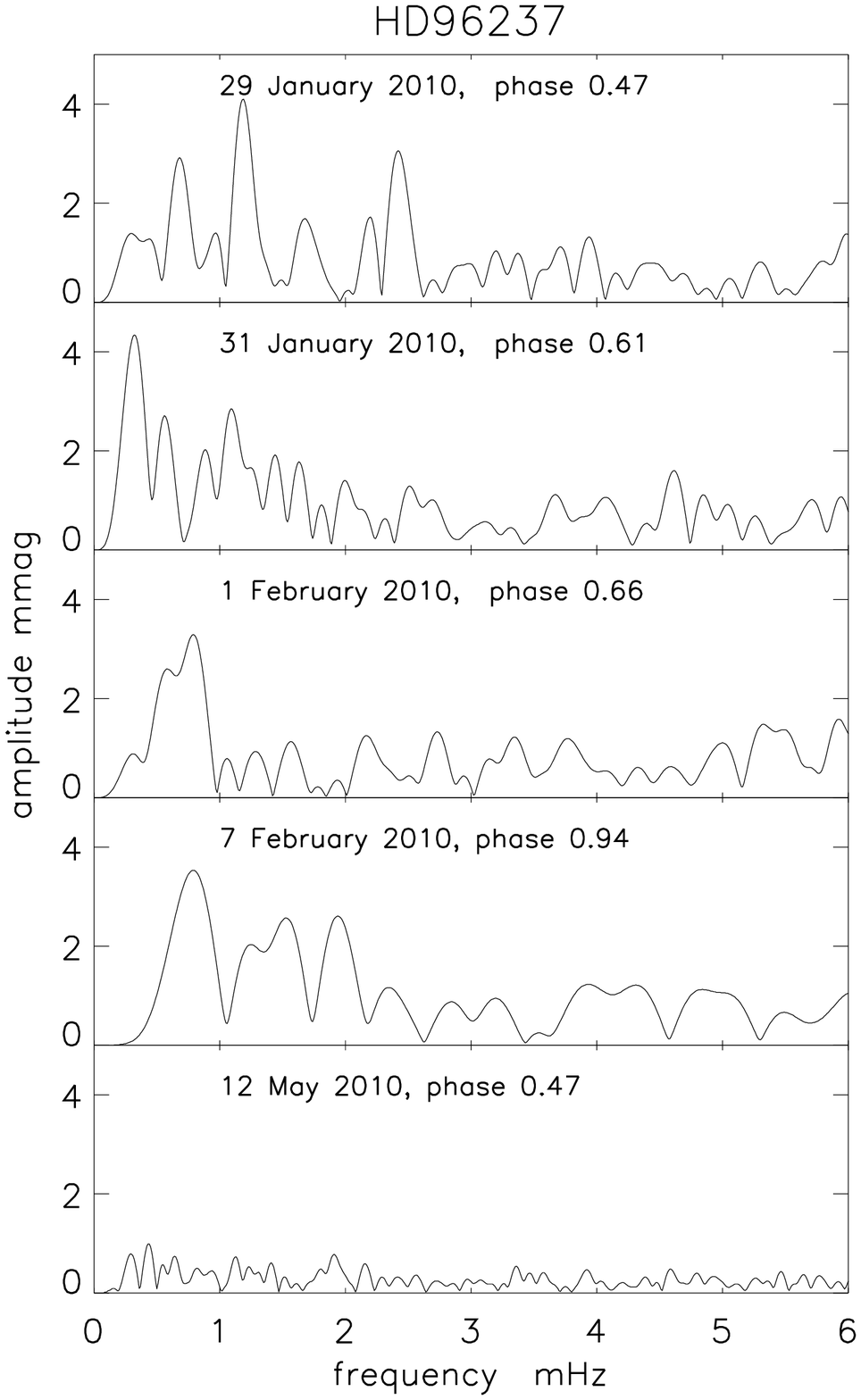}
\caption{\label{96237:phot} Amplitude spectra of the photometric observations for 
of HD\,96237. Extinction and sky transparency variations were removed using a 
second degree polynomial. The data are ordered in time from top to bottom. The 
highest peak in the upper panel has amplitude near 4\,mmag and frequency of 
1.19\.mHz, which is at the error limit of the frequency found for the 
spectroscopic data. The star may also show a harmonic with a frequency of 
2.42\,mHz. The third and fourth panels show peaks with the same frequency of 
0.79\,mHz. A similar frequency peak was detected in spectroscopic observations 
(see Fig.\,\ref{96237:ampspec}). It is not clear weather this peak is real; only 
one other roAp star, HD\,116114, with a similar low frequency has previously been 
found \citep{Elkin05}.}
\end{figure}

Photometric observations were obtained at SAAO for additional testing of 
pulsations in HD\,96237. The frequency analysis of the photometric data listed in 
Table~1 is presented in Fig.\,\ref{96237:phot}. The upper panel of this figure 
supports the pulsation period found by spectroscopy , but other photometric 
observations shown in the bottom four panels and by \cite{Nelson93} do not detect 
pulsation in this star.

This can be understood in terms of rotational modulation in an oblique pulsator, 
and/or beating of multiple frequencies. The pulsation amplitudes in many roAp 
stars are modulated with the rotation period of the star. Thus some of the 
photometric observations may not have been in best aspect. To judge this the 
rotational period and ephemeris of HD\,96237 needs to be determined. From the All 
Sky Automated Survey (ASAS) and Hipparcos photometry, \citet{Freyhammer08b} found 
photometric variations with a period of 20.91\,d, which may be the rotational 
period. Using two seasons of data from the WASP (Wide Angle Search for Planets) 
Project  (\citealt{Pollacco06}) covering the intervals 2007 January 4 to 2007 June 
3 and 2008 January 5 to 2008 May 28 we obtained a similar period. Combining the 
WASP (passband from 400 to 700\,nm) and ASAS (V-band) photometry we find the following rotational ephemeris for the photometric maximum:

\begin{equation}
HJD(phot. max.) = 2454483.9285 + 20.91E
\label{eq:epoch}
\end{equation}

The rotational phases calculated from this ephemeris are also shown in 
Fig.\,\ref{96237:phot}. The bottom panel is for data observed in the same rotation 
period as the upper panel, but does not display any pulsations. The spectral 
observations presented at Fig.\,\ref{96237:ampspec} were also obtained at a 
similar rotation phase using the above ephemeris. Assuming our case for pulsation 
in this star to be good, we suggest two possible solutions: 1) the rotation period 
may be not 20.91\,d, but double that. This ambiguity happens for some peculiar 
stars (see for example \cite{Wade97}).  Longitudinal magnetic field measurements 
over the rotational period can resolve this; 2) the star may be multiperiodic. 
There is some hint in Fig.\,\ref{96237:phot} of a peak at 0.79\,mHz.  Additional 
observations are required to resolve these questions.

\section{Concluding remarks}

The chemically peculiar magnetic star HD\,96237 is an interesting object which has 
a very peculiar spectrum with significant spectral variability. The star 
demonstrates large overabundances of rare earth elements. A high resolution 
spectrum obtained with the ESO 2.2-m telescope and FEROS spectrograph resembles 
the spectrum of another highly peculiar star, HD\,101065. Abundances of 
Nd\,\textsc{ii} and Nd\,\textsc{iii} determined from this spectrum are even higher 
than in HD\,101065. Other rare earth elements also show large overabundances 
similar to those found in HD\,101065. Another spectrum of the star obtained with 
VLT UVES was significantly different from the FEROS spectrum with much less 
intense spectral lines of the rare earth elements (\citealt{Freyhammer08b}). The 
similarity of the spectra and chemical abundances of HD\,96237 to those of 
HD\,101065, which was the first detected rapidly oscillating Ap star, increases 
the interest of this star. It is not known what determines the pulsation amplitude 
of the roAp stars. While the principal periods of HD\,96237 and HD\,101065 are 
similar (13.6\,min and 12.1\,min, respectively), their photometric and 
spectroscopic radial velocity amplitudes are very different.

Our results for HD\,69013 demonstrate clearly that this is a new roAp star with a 
low pulsation amplitude. This star shows possible rotation modulation and may be a 
useful target to study pulsation behaviour over rotation period. 

\section{Acknowledgements}

DWK and VGE acknowledge support for this work from the Science and Technology 
Facilities Council (STFC) of the UK. {This research has made use of NASA's 
Astrophysics Data System and SIMBAD database, operated at CDS, Strasbourg, 
France.} This paper uses observations made at the South African Astronomical 
Observatory (SAAO).

{}

\end{document}